\documentclass[iop, twocolumn]{aastex6}
\usepackage{graphics}
\usepackage{subfigure}
\usepackage{amsmath}

\newcommand{\cs}{c_{\rm s}}

\newcommand{\vturb}{v_{\rm turb}}

\begin{document}

\title{Origin of Weak Turbulence in the Outer Regions of Protoplanetary Disks}

\author{Jacob B. Simon\altaffilmark{1,2,3},  Xue-Ning Bai\altaffilmark{4,5,6},  Kevin M. Flaherty\altaffilmark{7}, A. Meredith Hughes\altaffilmark{7}}

\altaffiltext{1}{JILA, University of Colorado and NIST, 440 UCB, Boulder, CO 80309-0440}
\altaffiltext{2}{Department of Space Studies, Southwest Research Institute, Boulder, CO 80302}
\altaffiltext{3}{Kavli Institute for Theoretical Physics, UC Santa Barbara, Santa Barbara, CA 93106}
\altaffiltext{4}{Harvard-Smithsonian Center for Astrophysics, 60 Garden St.,
MS-51, Cambridge, MA 02138}
\altaffiltext{5}{Institute for Advanced Study, Tsinghua University, Beijing 100084, China}
\altaffiltext{6}{Tsinghua Center for Astrophysics, Tsinghua University, Beijing 100084, China}
\altaffiltext{7}{Astronomy Department, Van Vleck Observatory, Wesleyan University, 96 Foss Hill Dr., Middletown, CT 06459}

\email{jbsimon.astro@gmail.com}

\begin{abstract}
The mechanism behind angular momentum transport in protoplanetary disks, and whether this transport is turbulent in nature, is a fundamental issue in planet formation studies. Recent ALMA observations have suggested that turbulent velocities in the outer regions of these disks are less than $\sim 0.05$--$0.1\cs$, contradicting theoretical predictions of turbulence
driven by the magnetorotational instability (MRI).  These observations have generally been interpreted to be consistent with a large-scale laminar magnetic wind driving accretion. Here, we carry out local, shearing box simulations with varying ionization levels and background magnetic field strengths in order to determine which parameters produce results consistent with observations. We find that even when the background magnetic field launches a strong largely laminar wind, significant turbulence persists and is driven by localized regions of vertical magnetic field (the result of zonal flows) that are unstable to the MRI.   The only conditions for which we find turbulent velocities below the observational limits are
weak background magnetic fields and ionization levels well below that usually assumed in theoretical studies.   We interpret these findings within the context of a preliminary model in which a large scale magnetic field, confined to the inner disk, hinders ionizing sources from reaching large radial distances, e.g., through a sufficiently dense wind. Thus, in addition to such a wind, this model predicts that for disks with weakly turbulent outer regions, the outer disk will have significantly reduced ionization levels compared to standard models and will harbor only a weak vertical magnetic field.
\end{abstract}


\section{Introduction}

A long standing question in accretion disk theory is how exactly angular momentum is removed from orbiting gas, allowing it to accrete onto the central object.  While this issue is universal to all accretion disks, understanding angular momentum transport in protoplanetary disks is particularly crucial to understanding planet formation. Indeed, whether angular momentum is transported via turbulence or some other process has significant implications for a range of planet formation stages, including the settling and growth of dust grains \citep{fromang06a,youdin07b,birnstiel10}, the concentration of particles \citep{cuzzi08,johansen09a,simon14}, and the migration of planetary bodies \citep{nelson04,lubow11,baruteau11,paardekooper11}.

For many years, it has been theorized that turbulence driven by the magnetorotational instability \cite[MRI;][]{balbus98} is responsible for angular momentum transport in such disks; angular momentum is redistributed radially via correlated turbulent fluctuations in the radial and azimuthal components of both magnetic fields and gas velocities.  Recently, however, protoplanetary disk theory has undergone a paradigm shift.   In particular, studies including non-ideal magnetohydrodynamic (MHD) effects other than Ohmic resistivity (as conventionally considered, e.g., \citealt{gammie96}) resulting from low ionization fractions have found that MRI-driven turbulence is largely reduced at the mid-plane in the outer disk \citep{simon13a,simon13b}, whereas it remains active in the upper layers due to ionization from FUV photons \cite[e.g.,][]{perez-becker11b,simon13b}. In the inner disk, MRI-driven turbulence can be either reduced significantly \citep{simon15b} or quenched altogether \cite[e.g.,][]{bai13b,lesur14}, depending on the precise levels of ionization \citep{simon15b}. Furthermore, angular momentum can be additionally removed vertically through a magnetic wind \citep{salmeron07,suzuki09,bethune17,bai17a}, akin to the Blandford-Payne process \citep{blandford82} and radially through laminar torques \cite[e.g.,][]{lesur14}.

With the advent of ALMA, we can now directly test these ideas with high spatial and spectral resolution data.  In our recent work, (\citealt{flaherty15} and \citealt{flaherty17}, hereafter F15 and F17, respectively), we analyzed ALMA data using Markov-chain Monte Carlo techniques to put a strong upper limit on turbulent velocities in HD163296.  We found that $\vturb \lesssim 0.05\cs$ throughout the entire disk column in the regions most easily resolved with ALMA ($\sim 30$--100 AU and beyond), corresponding to a Shakura-Sunyaev $\alpha < 10^{-3}$. While these constraints are consistent with theoretical expectations from numerical simulations for the {\it mid-plane} region, where ambipolar diffusion damps MRI-driven turbulence (\citealt{simon15a}, hereafter S15), they are in fact {\it inconsistent} with predictions that $\vturb \sim 0.1$--1$\cs$ in the upper layers of the disk, where FUV ionization is sufficiently strong for the MRI to operate \cite[e.g.,][]{perez-becker11b,simon13b}.  Furthermore, such a low $\alpha$ is inconsistent with the observed accretion rates (assuming a steady-state accretion flow driven by turbulence) onto the central star, $\dot{M} \sim 10^{-7} M_{\odot}/{\rm yr}$ \citep{mendigutia13}. 

In addition, other observations and analyses have pointed toward weak turbulence in the outer regions of protoplanetary disks.  We have recently found that TW Hya also exhibits weak turbulence at large radial distances, with $\vturb \lesssim 0.08\cs$ \citep{flaherty18}. Furthermore, modeling of the dust in the HL Tau system by \cite{okuzumi16} and \cite{pinte16} has suggested that turbulence in this system must be weak (\citealt{pinte16} estimates $\alpha \sim 10^{-4}$) in order for the rings to achieve their observed structure.  If indeed turbulent velocities are as weak as these studies suggest, the inconsistencies between observations and theoretical predictions call into question the idea of turbulence-driven angular momentum transport.  The question that then remains is: can magnetic winds drive accretion in a laminar fashion?

In this paper, we carry out a series of local (i.e., small co-rotating disk patch) simulations to explore accretion driven both by magnetically launched winds and turbulence, focusing in particular on the observations constraints from the HD163296 disk.  We find that, only in the limit of weak ionization {\it and} a weak vertical magnetic field threading the disk can we reproduce the small turbulent velocities as measured by F15 and F17.  In particular, even when there is a substantial magnetic wind, the gas flow is not laminar, but instead quite turbulent. Such turbulence, if present in HD163296, would have been detected in the data from F15 and F17.  These results suggest that the outer disk is only very weakly accreting, if at all.  One possible solution to the low-turbulence issue (as we discuss in detail below) is that ionizing radiation is being blocked from reaching the outer disk, where only a very weak magnetic field is present.  Along these lines, we construct a preliminary model in which a large scale magnetic field, mostly confined to small radial distances, can block ionizing radiation from the outer disk.  We use this model to explain current observational constraints and make testable predictions for future observations, both of HD163296 and other systems.

The outline of our paper is as follows.  In Section~\ref{method}, we describe the details of our calculations, including the algorithm and parameters chosen.  In Section~\ref{results}, we present our main results, and we follow up with an in-depth analysis of these results in Section~\ref{origin}.  We then discuss the implications of our study and our preliminary model to explain the observations in Section~\ref{discussion}.  Finally, we wrap up with our main conclusions in Section~\ref{conclusions}.

\begin{widetext}
\begin{deluxetable*}{l|ccccccccc}
\tabletypesize{\small}
\tablewidth{0pc}
\tablecaption{Numerical Simulations\label{tbl:sims}}
\tablehead{
\colhead{Label}&
\colhead{$R_0$}&
\colhead{$\beta_0$}&
\colhead{FUV?}&
\colhead{X-ray?}&
\colhead{CR?}&
\colhead{$\left\langle\frac{\delta v}{\cs}\right\rangle_{\mathcal{G}, z \ge 2H}$}&
\colhead{$\alpha$} \\
\colhead{ }&
\colhead{(AU)}&
\colhead{ }&
\colhead{ }&
\colhead{ }&
\colhead{ }&
\colhead{ }&
\colhead{ }&
\colhead{ } } 
\startdata
R100-B3p-FXC-6H & 100 & $10^3$ & Yes & Yes & Yes & 0.39 & $2.7\times10^{-2}$ \\
R100-B4p-FXC & 100 & $10^4$ & Yes & Yes & Yes & 0.25 & $9.7\times10^{-3}$ \\
R100-B4n-FXC & 100 & $-10^4$ & Yes & Yes & Yes & 0.45 & $7.4\times10^{-3}$ \\
R100-B5p-FXC & 100 & $10^5$ & Yes & Yes & Yes & 0.73 & $2.7\times10^{-3}$ \\
R100-B6p-FXC & 100 & $10^6$ & Yes & Yes & Yes & 0.61 & $8.4\times10^{-4}$ \\
R100-B7p-FXC & 100 & $10^7$ & Yes & Yes & Yes & 0.38 & $2.3\times10^{-4}$ \\
R100-B3p-XC-5.25H & 100 & $10^3$ & No & Yes & Yes & 0.21 & $1.1\times10^{-2}$ \\
R100-B4p-XC-6H & 100 & $10^4$ & No & Yes & Yes & 0.30 & $2.9\times10^{-3}$ \\
R100-B4n-XC-6H & 100 & $-10^4$ & No & Yes & Yes & 0.13 & $2.4\times10^{-4}$ \\
R100-B5p-XC-6H & 100 & $10^5$ & No & Yes & Yes & 0.14 & $7.0\times10^{-4}$ \\
R100-B6p-XC-6H & 100 & $10^6$ & No & Yes & Yes & 0.09 & $2.3\times10^{-4}$ \\
R100-B7p-XC-6.5H & 100 & $10^7$ & No & Yes & Yes & 0.04 & $5.9\times10^{-5}$ \\
R100-B4p-X-5.5H & 100 & $10^4$ & No & Yes & No & 0.29 & $1.7\times10^{-3}$ \\
R100-B4p-C-5.5H & 100 & $10^4$ & No & No & Yes & 0.20 & $2.3\times10^{-2}$ \\
R100-B4p-CWeak-5.25H & 100 & $10^4$ & No & No & Very weak & 0.08 &$1.8\times10^{-3}$  \\
R100-B5p-CWeak-5.25H & 100 & $10^4$ & No & No & Very weak & 0.02 &  $5.7\times10^{-4}$\\
R200-B4p-XC-5.25H & 200 & $10^4$ & No & Yes & Yes & 0.18 & $1.4\times10^{-3}$ \\
\enddata
\end{deluxetable*}
\end{widetext}

\section{Method}
\label{method} 

To carry out our investigation, we simulate the evolution of a small accretion disk patch using the
local, shearing-box approximation \citep{hawley95a}, including vertical stratification and all three non-ideal effects (Ohmic
and ambipolar diffusion, and the Hall effect). The local simulations are centered at 100AU in a model for the disk around HD163296 \citep{rosenfeld13}, following the approach of S15.
 We employ the {\sc Athena} code \citep{stone08} with the implementation of ambipolar diffusion and the Hall effect described by \cite{bai11b} and \cite{bai14a}, respectively.  Furthermore, we use {\sc Athena} in a configuration identical to that in S15. Following that work, we obtain the diffusivities using a look-up table calculated via a chemical network as described further in \cite{bai13b}, \cite{bai13c}, and \cite{bai14a}. We include cosmic ray ionization at a rate  $\xi_{\rm cr}=10^{-17}\exp{\left[-\Sigma/(96{\rm~g~cm}^{-2})\right]}$s$^{-1}$ \citep{umebayashi81}, though in two of our simulations, we set the pre-factor of this quantity to $10^{-19} {\rm s}^{-1}$ (see description below).  We also include X-rays as in S15 with an X-ray luminosity $L_X= 4\times10^{29}$ erg s$^{-1}$ and X-ray temperature corresponding to 1 keV, specifically appropriate to HD163296 \citep{swartz05,gunther09}. Additionally, we include $^{26}$Al decay at an ionization rate $10^{-19} {\rm s}^{-1}$.

 In the surface layer, we further include the effect of far UV ionization based on calculations from \cite{perez-becker11b}. The FUV ionization is assumed to have a constant penetration depth of $\Sigma_{\rm FUV} = 0.01$~g~cm$^{-2}$.   In these surface layers, we calculate the ionization rate by setting the 
 ambipolar diffusion Elsasser number ${\rm Am} = \gamma\rho_i/\Omega$, where $\gamma$ is the coefficient of momentum transfer for ion-neutral collisions, $\rho_i$ is the mass density of ions, and $\Omega$ is the orbital frequency, and then calculating the diffusivity from Am.  Compared with S15, we use a modified version of Am in the FUV-ionized layer more appropriate to the structure of HD163296,

\begin{equation}
\label{Am_FUV}
{\rm Am_{\rm FUV}} = 2.4\times10^{16}
\rho_{0,{\rm mid}}\bigg(\frac{f}{10^{-5}}\bigg)\left(\frac{r}{{\rm 1 AU}}\right)^{3/2},
\end{equation}

\noindent
where $\rho_{0,{\rm mid}}$ is the mid-plane gas density in cgs units, $f$ is the ionization fraction, and $r$ is the radial coordinate away from the star.
In calculating the diffusivities, we assume no grains are present since S15 demonstrated that the inclusion of 0.1 $\mu$m grains made very little difference to the amplitude of turbulence. We further discuss the neglect of dust grains in Section~\ref{discussion}.

The initial conditions are identical to the simulations of S15. 
The standard domain size is $4H \times 8H \times 8H$ in $(x,y,z)$, 
where the vertical scale height $H = \sqrt{2} c_s / \Omega$, with $c_s$ and $\Omega$ being the sound speed and orbital frequency, respectively.  We decrease the vertical extent of the box for
some of our simulations in order to reduce the overhead associated with small time steps set by low-density gas.  We employ a resolution of 36 grid zones per $H$ in all calculations.  All simulations assume an isothermal equation of state. We parameterize the net vertical flux threading the simulation volume in terms of $\beta_0 \equiv {\rm sign}(B_{z,0})2 P_0/B_{z,0}^2$, where $P_0$ is the initial mid-plane gas pressure and $B_{z,0}$ is the strength of the background vertical field. Note that due to the dependence of the Hall effect on the sign of this vertical field \cite[e.g.,][]{kunz13,lesur14,bai15,simon15b}, $\beta_0$ includes this sign dependence.   We run each calculation out so that a statistically steady state is reached.  

In what follows, we vary $\beta_0$, the radial location of the shearing box
$R_0$, and the strength of various ionization sources. These simulations are summarized in Table~\ref{tbl:sims}. 
The label of each simulation is R\#-B\#(p,n)-FXC, where the first \# is the radial location of the shearing box in units of AU, the second $\#$ is the
exponent of $\beta_0$ (e.g., 3 for $\beta_0 = 10^3$), (p,n) refers to a positive or negative value for $\beta_0$, and the inclusion of ``F", ``X", ``C" denotes the inclusion of FUV, X-rays, and cosmic rays, respectively (in our simulations, $^{26}$Al decay is always present as an additional ionization source).  As mentioned above, some of the simulations were run in a shorter domain in order to ensure that the time step (which is controlled by the lowest density gas near the vertical boundaries) remains reasonable.  For these calculations, we append the label with $\#H$ where \# is the total number of vertical scale heights spanned by the domain.  Note that two additional runs were done with cosmic rays and $^{26}$Al decay only, but with the cosmic ray ionization rate set to $10^{-19} {\rm s}^{-1}$; these runs are labelled with ``CWeak".

\section{Turbulent Velocities}
\label{results}

\begin{figure}[t!]
\begin{center}
\includegraphics[width=0.5\textwidth,angle=0]{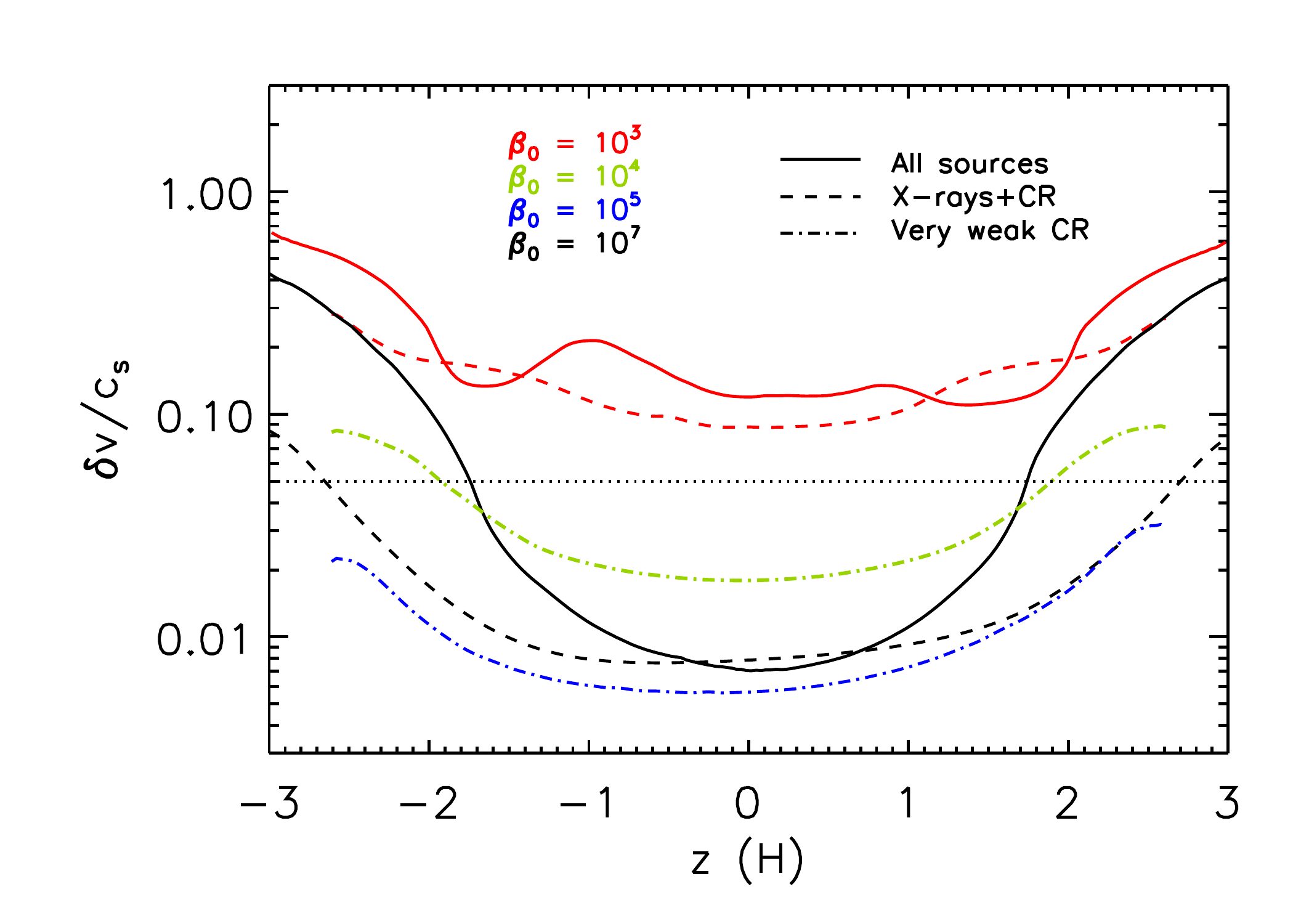}
\end{center}
\caption{Horizontally and temporally averaged $\delta v/\cs$ versus $z$ for magnetic field strengths characterized by $\beta_0 = 10^3$ (red), $10^4$ (green), $10^5$ (blue), and $10^7$ (black), with all sources of ionization (solid lines), X-rays and cosmic rays only (dashed), and only cosmic rays at a reduced flux (dot-dashed). The horizontal dotted line represents the approximate upper limit on the turbulent velocity $\delta v < 0.05\cs$ from F15 and F17.  The simulations that fall below this line (or are somewhat marginal) are for weak magnetic fields ($\beta_0 > 10^5$) and reduced ionization.
}
\label{vt_beta}
\end{figure}

\begin{figure}[t!]
\begin{center}
\includegraphics[width=0.5\textwidth,angle=0]{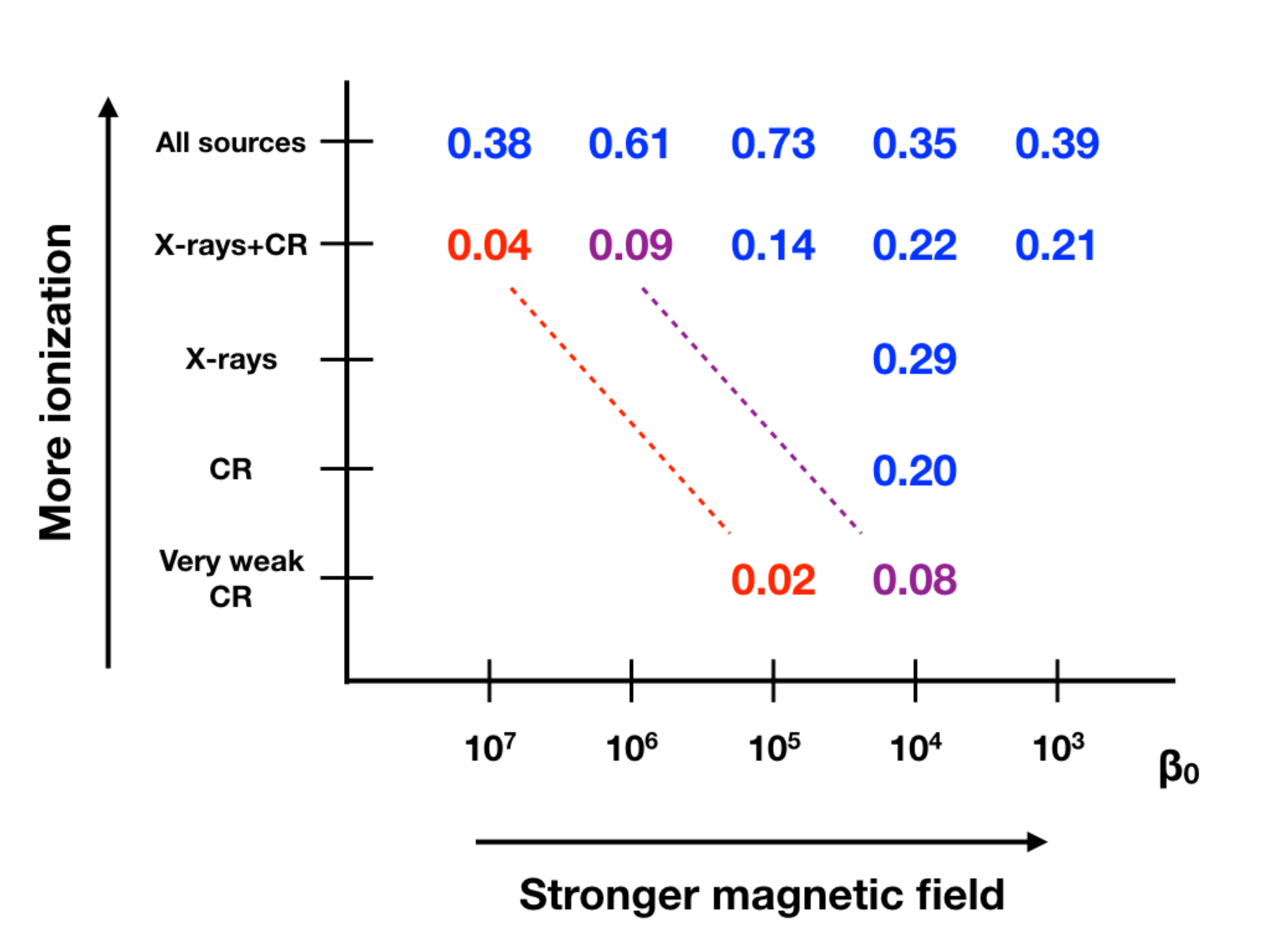}
\end{center}
\caption{Spatial (geometric) and time average of $\delta v/\cs$ for $|z| \ge 2H$ as a function of $\beta_0$ and ionization level.  Numbers in blue are well above the upper limit of $0.05\cs$ (from F17),
whereas red numbers are below this limit.  The purple numbers correspond to $\delta v < 0.1\cs$ and may be considered marginal.  The dashed lines suggest a region of parameter space in which a weaker magnetic field can be traded off for stronger ionization in order to maintain low turbulence values.
}
\label{xe_beta}
\end{figure}

Following the arguments in S15, we calculate the turbulent velocity, $\delta v$, by subtracting the shear flow, the laminar wind component, and the systematic increase or decrease in azimuthal velocity due to zonal flows and then adding the remaining velocity components in quadrature. In what follows we time-average the turbulent velocity over a period in which various flow quantities (e.g., Maxwell stress) are statistically constant in time. Figure~\ref{vt_beta} shows this velocity, normalized to $\cs$, versus $z$ for a range of background magnetic field strengths, ranging from $\beta_0 = 10^7$ to $\beta_0 = 10^3$.   The complete run list, along with a geometrically averaged (denoted by $\mathcal{G}$) turbulent velocity (for $|z| > 2H$) is shown in Table~\ref{tbl:sims} and is demonstrated pictorially in Fig.~\ref{xe_beta}.  We also include the standard \cite{shakura73} $\alpha$ value in the table.

It is clear from both the figures and the table, that for a large range in magnetic field strengths, $\beta_0 = 10^3$--$10^7$, there are significant turbulent velocities at large $|z|$, well above the upper limit put on this velocity by F15 and F17, represented approximately by the dotted line in Fig.~\ref{vt_beta}.  As $\beta_0$ decreases, the magnetic field structure tends towards being more laminar as shown by \cite{simon13b} and verified here.  Thus, for different magnetic field configurations (largely laminar, largely turbulent, or some combination thereof), we still find velocities above observational constraints at large $|z|$.  For fields weaker than $\beta_0 \sim 10^3$--$10^4$, the turbulence in the mid-plane is well below this upper limit and thus consistent with these observations.  Even the simulations in which only one source of ionization (in addition to $^{26}$Al decay) is included at the strength used previously (S15) show appreciable turbulence at large $|z|$ for all but the weakest vertical fields.

We have varied other parameters including magnetic field polarity, $R_0$, and the strength of ionization sources.  In nearly all cases, we find strong turbulence with $\delta v \sim 0.1$--1$\cs$ at $|z| \sim 2$--$4H$.  Based on the data in Table~\ref{tbl:sims} and Fig.~\ref{xe_beta}, there is a trade-off between field strength and ionization level; for higher ionization, a weaker background field is necessary to produce low turbulence values.

\begin{figure*}[ht!]
\begin{center}
\includegraphics[width=0.47\textwidth,angle=0]{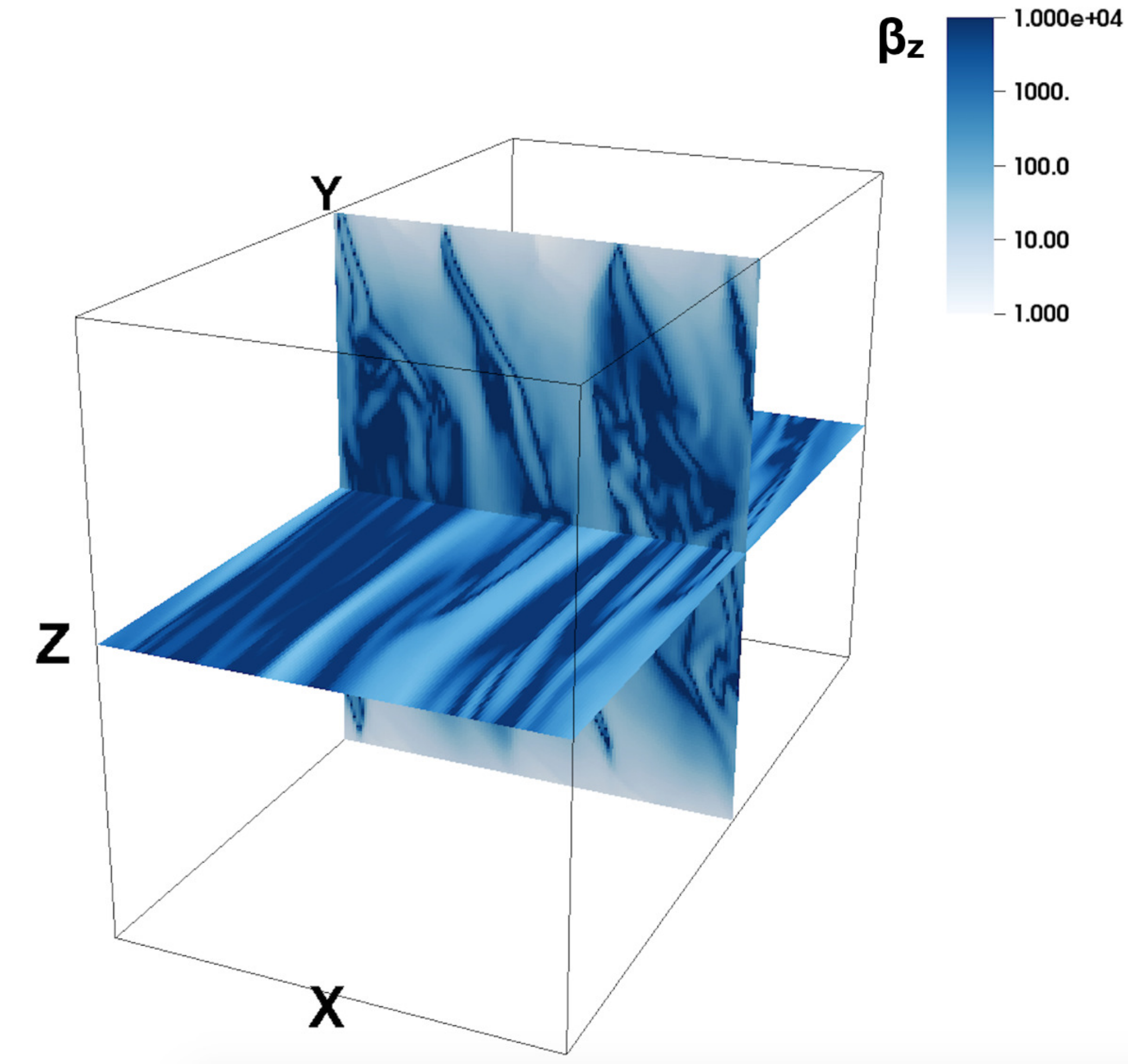}
\includegraphics[width=0.47\textwidth,angle=0]{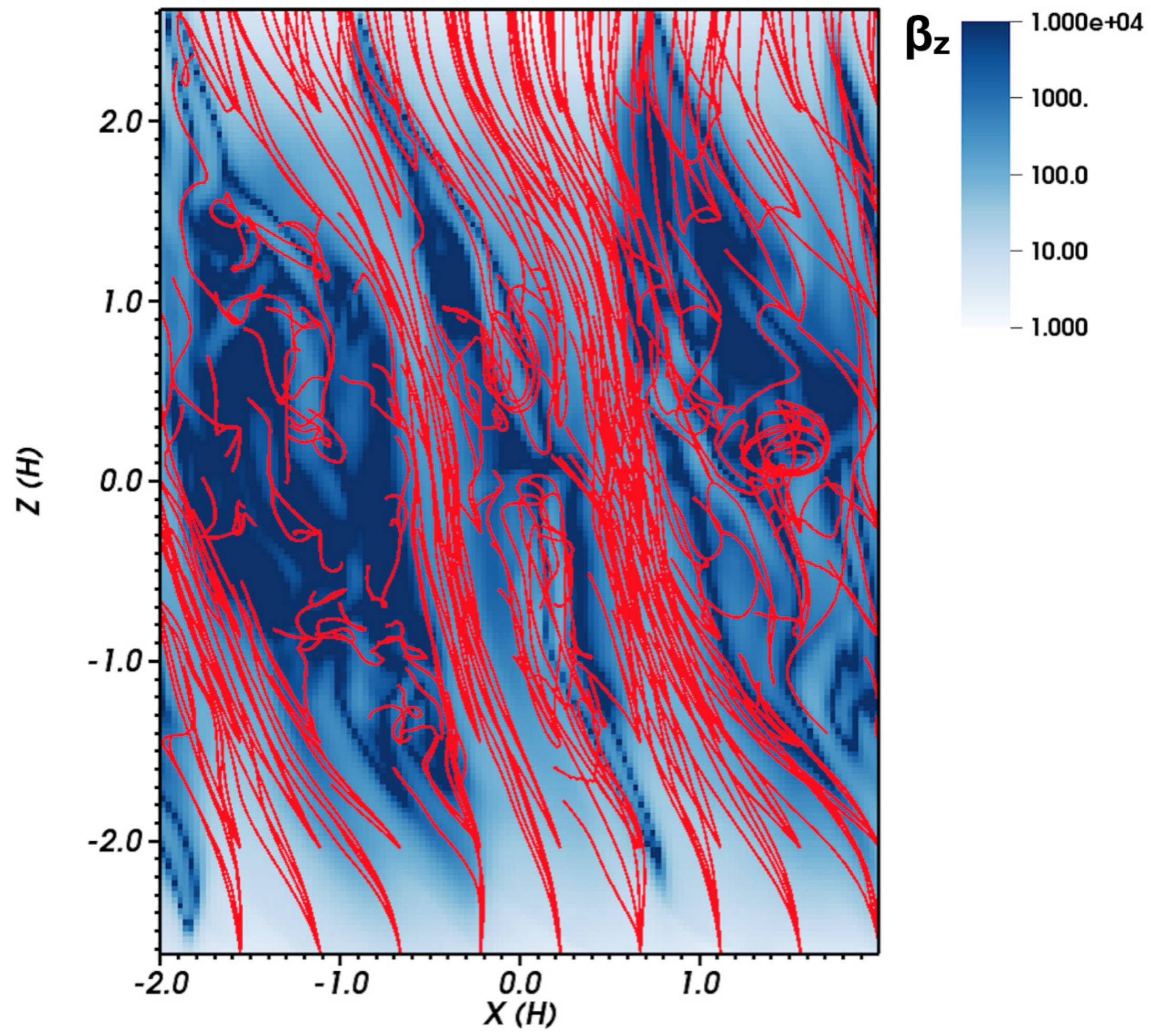}
\end{center}
\caption{
Spatial distribution of the local $\beta_z \equiv 2 P/B_z^2$ for a snapshot at orbit 30 in R100-B3p-XC-5.25H. The left panel shows
slices along the mid-plane and $y = 0$ in the simulation domain.  The right panel shows the $y = 0$ slice with magnetic field
lines superimposed (red lines).  There are large scale, mostly axisymmetric, spatial variations in $\beta_z$. In large $\beta_z$ regions, the magnetic field is tangled due to turbulence, whereas for $\beta < 10^3$, the field is relatively laminar.}
\label{beta}
\end{figure*}
Note that for $\beta_0 \ge 10^6$, the MRI is largely under-resolved near the disk mid-plane.  However, since the strength of MRI turbulence decreases with increasing $\beta_0$ \citep{bai11a}, the trend of the mid-plane $\delta v$ values with $\beta_0$ suggests that $\delta v$ will be quite small there, and will not likely strongly affect the turbulence in the upper layers.

Furthermore, due to a restrictively small time step, we could not run stratified simulations with $\beta_0 < 10^3$ and Am $\ll$ 1. However, with a strong net field (e.g., $\beta_0 \leq 10^2$), one can conceive of two regimes. First, for Am sufficiently low to suppress/damp the MRI but still sufficiently large for there to be some coupling between gas and magnetic field, a strong field would drive a substantial wind, resulting in accretion rates well in excess of the observed rate \cite[e.g.,][]{simon13b}.  Second, for extremely weak ionization (Am$\rightarrow0$), the field and gas are completely decoupled, and hence no accretion. An intermediate regime in which the coupling and field strength are just at the right level to give the right accretion rate may exist, but as we discuss in Section~\ref{cweak}, even here, one expects to see turbulent gas velocities.

\section{Origin of the Turbulence}
\label{origin}

Our results clearly show that even in the presence of a predominantly wind-driven accretion flow (i.e., the magnetic field structure appears relatively laminar), there are still significant turbulent velocities.   This result is surprising in light of recent results demonstrating that for a sufficiently strong magnetic field, ambipolar diffusion quenches the turbulence (see Fig. 16 in \citealt{bai11a}). To further understand these results, we now focus on determining the origin of this turbulence.  

\newpage
\subsection{Local MRI Turbulence}

We have examined the spatial structure of the magnetic field and $\beta_z \equiv 2 P/B_z^2$, where $\beta_z$ is defined locally and not as the background value.  Figure~\ref{beta} shows $\beta_z$ for a snapshot at orbit 30 in R100-B3p-XC-5.25H, for which $\beta_0 = 10^3$. It appears that the magnetic flux has been redistributed into nearly axisymmetric structures.  With this redistribution of flux, large regions of the box have $\beta \geq 10^4$, sufficiently large for the MRI to be active in those regions, based on Fig. 16 of \cite{bai11a}.\footnote{Technically, the condition imposed by \cite{bai11a} includes the toroidal component of the magnetic field as well as the vertical.  However, recent work by Gole \& Simon, submitted, have shown that for Am values consistent with those used in our work, a significant fraction of the toroidal field within the domain is sufficiently weak so as to allow turbulence.}  That these regions are MRI active is supported by the right panel of the figure, in which a meridional slice of the domain is shown with magnetic field lines over-plotted.  In the regions of low $\beta_z$, the field lines are laminar, whereas they become quite tangled when $\beta_z \gtrsim 10^4$. 

Examining our remaining calculations, we find that for $10^3 \le \beta_0 \le10^4$ (with the exception of the run R100-B4p-CWeak-5.25H; see below) zonal flows are established near the mid-plane, which lead to regions of magnetic flux that are sufficiently weak to drive the MRI. In all of these cases, while the turbulence apparently originates near the mid-plane, velocity fluctuations are likely amplified as they propagate towards the lower density regions at large $|z|$ \citep{simon11b}.  

\begin{figure*}[ht!]
\begin{center}
\includegraphics[width=1\textwidth,angle=0]{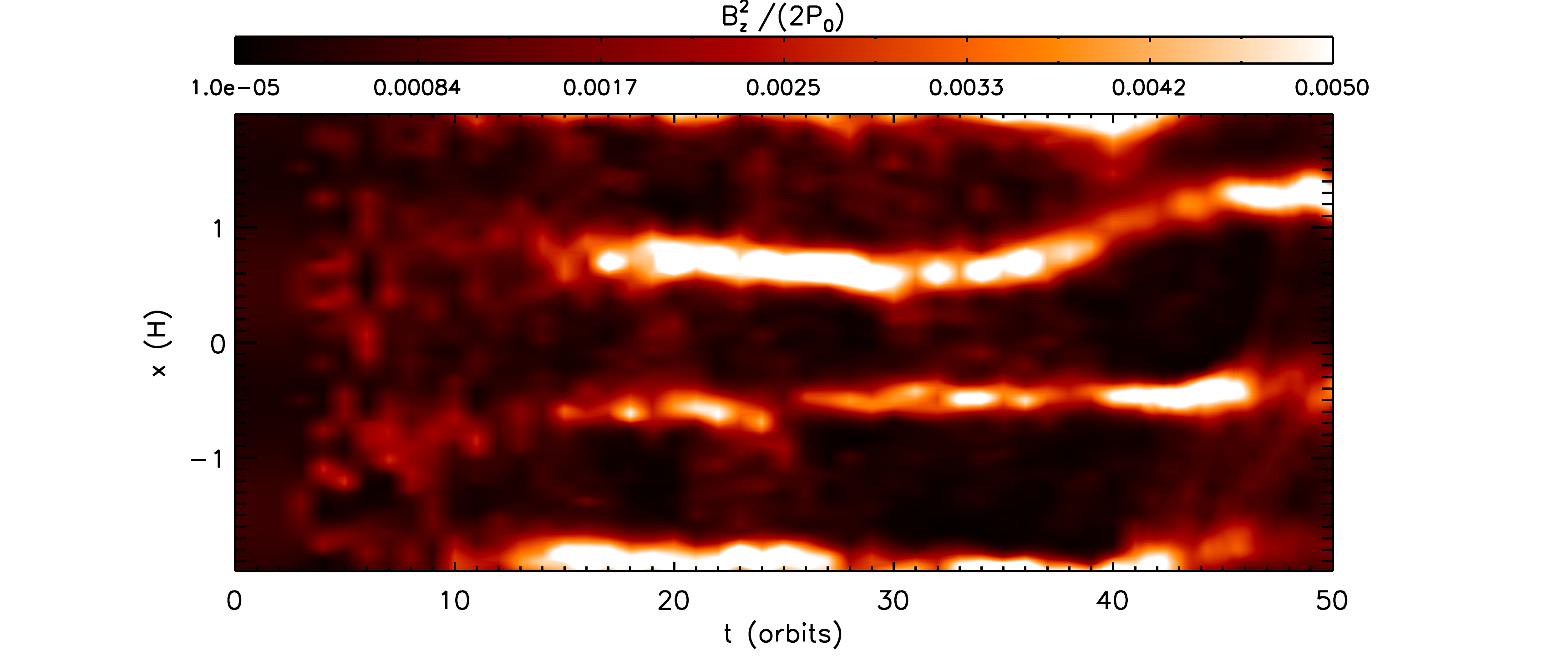}
\end{center}
\caption{Space-time diagram ($t,x$) of the vertical magnetic energy normalized to the initial mid-plane gas pressure for the simulation R100-B3p-XC-5.25H.  The average of the magnetic energy was taken over all $y$ and over $|z| < 0.5H$.  The emergence of very narrow bands of large magnetic flux emerge after $\sim 15$ orbits.  These structures have been seen in the simulations of \cite{bai15} and coincide with the strong magnetic flux regions shown in Fig.~\ref{beta}.
}
\label{sttx_bez}
\end{figure*}

For simulations with $10^5 \le \beta_0 \le 10^7$ (with the exception of the run R100-B5p-CWeak-5.25H; see below), the background magnetic field is sufficiently weak that there is MRI-driven turbulence throughout a large portion of the domain (again, based on the condition that weaker fields lead to turbulence; \citealt{bai11a}).  It is worth pointing out again that for runs with $\beta_0 \ge 10^5$, the MRI is largely under-resolved near the mid-plane.  Thus, higher resolution simulations will be required to determine the precise amplitude and structure of the turbulence in these cases. However, as we discussed above, for sufficiently weak magnetic field, the turbulent velocities induced by the MRI will likely be small enough that they will not appreciably change our results. 

Beyond the MRI, other instabilities that could potentially generate the turbulence seen in our simulations are the Hall-shear instability \cite[HSI;][]{kunz08} and the ambipolar diffusion shear instability \cite[ADSI;][]{kunz08}. We have run one additional simulation in which we neglect the Hall effect and set Am = 1 everywhere.   We still see significant turbulence with a similar structure to that seen in the runs with all three non-ideal terms.  This result suggests that the HSI, while potentially present in our simulations, does not play the dominant role in driving turbulence in these simulations.

We cannot say for certain that the ADSI is absent in our simulations, as to our knowledge, no well-tested numerical study of the non-linear ADSI has been done.  However, as suggested by the work of \cite{pandey12}, this instability can be considered as an extension of the MRI in the strong ambipolar diffusion dominated regime.

Taken together, these considerations strongly suggest that the MRI is largely responsible for generating turbulence in our simulations, even in the case of a relatively strong magnetic field, and that this turbulence originates from regions of relatively small vertical magnetic flux. Furthermore, in some of our simulations, a current sheet persists near the disk mid-plane, which is the result of large scale toroidal fields of opposite polarities coming into contact.  These current sheets generate velocity fluctuations, which likely become amplified as they move towards lower density regions \cite[see][]{simon11b}.  In summary, even so-called laminar wind flows are not entirely laminar.

\begin{figure}[t!]
\begin{center}
\includegraphics[width=0.5\textwidth,angle=0]{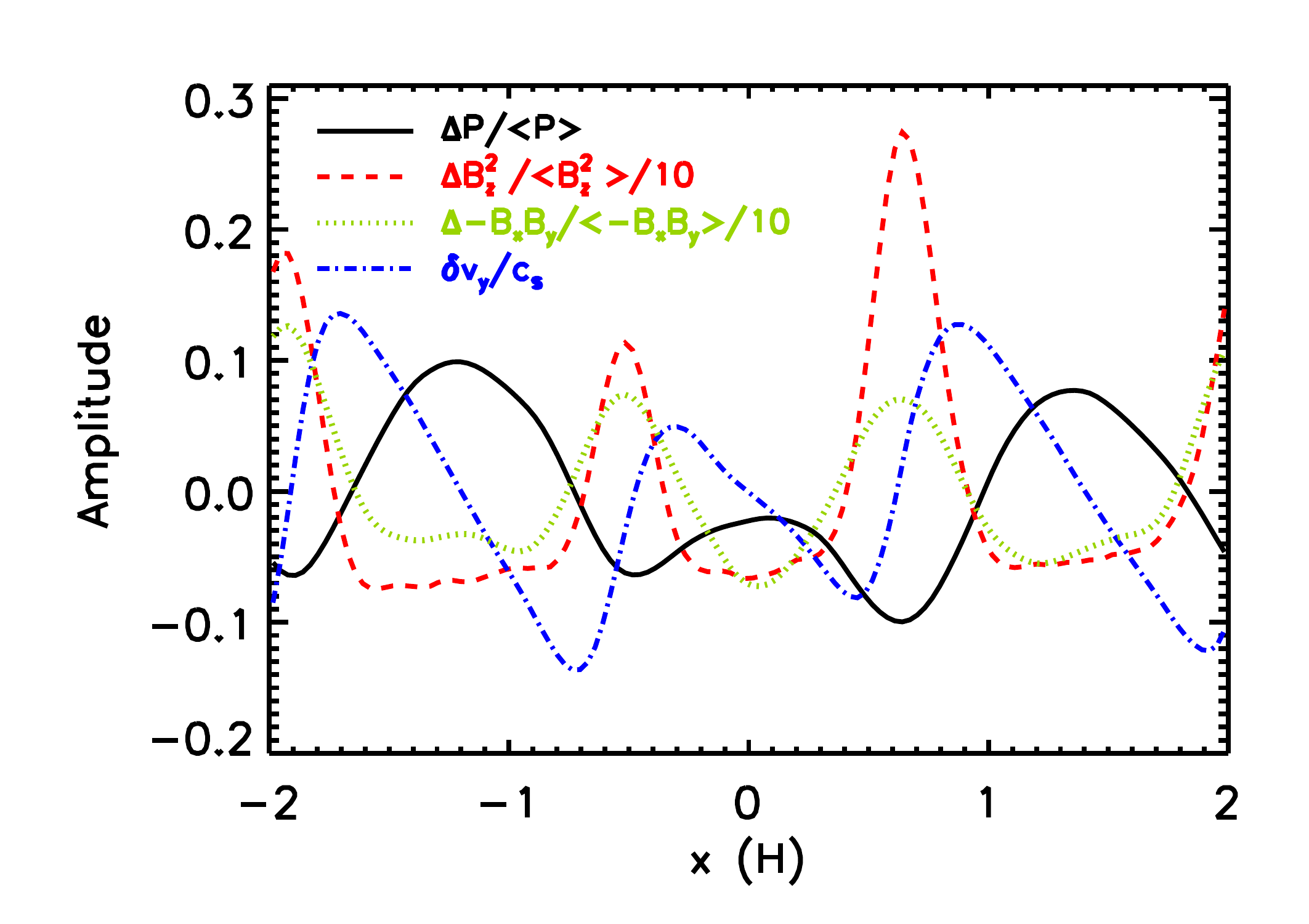}
\end{center}
\caption{Radial profile of gas pressure (black, solid), vertical magnetic energy (red, dashed; this quantity reduced by a factor 10 for visibility), Maxwell stress (green, dotted; this quantity reduced by a factor of 10 for visibility), and azimuthal velocity (blue, dot-dashed) for the simulation R100-B3p-XC-5.25H. The relative perturbation amplitude, $\Delta Q/\langle Q\rangle = (Q - \langle Q\rangle)/\langle Q\rangle$, is shown for the magnetic energy ($Q = B_z^2$), Maxwell stress ($Q = -B_xB_y$), and gas pressure ($Q = P$), where each quantity has been averaged over all of $y$, for $|z| < 0.5H$, and from orbit 20 to 35. The angled brackets denote an average over $x$.  The azimuthal velocity (here normalized by the sound speed) is similarly averaged over $y,z$ and time.  The perturbations in vertical magnetic energy are out of phase with the gas pressure perturbations by $\pi$ and similarly out of phase with the azimuthal velocity by $\sim \pi/2$; these properties are consistent with zonal flows.
}
\label{st_profile}
\end{figure}

\subsection{Zonal Flows}

We have analyzed the axisymmetric structures seen in R100-B3p-XC-5.25H in order to better characterize and understand their properties.  Figure~\ref{sttx_bez} shows the space-time diagram of the vertical magnetic energy averaged over all $y$ and for $|z| < 0.5H$.  As the figure shows, after $\sim15$ orbits, very narrow bundles of vertical magnetic flux emerge and persist until the end of the simulation. The radial width of these structures (as measured at the base; see Fig.~\ref{st_profile}) is approximately $\sim 0.5H$--$H$, consistent with the zonal flows seen in \cite{bai14b}, and especially \cite{bai15}, where the very narrow width of the magnetic flux bundles here roughly match that seen in that work.  \cite{bai14b} explained the emergence of these zonal flows with a phenomenological model.  Given the differences between the simulations in that work (e.g., unstratified, no Hall effect) and ours, it is beyond the scope of our work to match precisely the properties of the zonal flows we see here compared to these previous works.  However, some generic properties should persist if these structures are indeed zonal flows, which we now test.

Figure~\ref{st_profile} plots the relative perturbation amplitude, $\Delta Q/\langle Q\rangle = (Q - \langle Q\rangle)/\langle Q\rangle$ for some quantity $Q$ which has already been averaged over all $y$, over $|z| < 0.5H$, and in time from orbits 20 to 35, during which the radial movement of the magnetic flux bundles is small (see Fig.~\ref{sttx_bez}). The angled brackets denote an average over $x$.  Here, we plot this relative perturbation for the vertical magnetic energy, Maxwell stress, and the gas pressure.  We also plot the perturbed (i.e., Keplerian-shear-subtracted) azimuthal velocity normalized by the sound speed $\delta v_y/\cs$, time-averaged over the same period of time as the other quantities.  Note that we had to reduce the vertical magnetic energy and Maxwell stress by a factor of 10 for it to fit on the same plot.   Consistent with previous zonal flow models \cite[e.g.,][]{johansen09a,bai14b}, the magnetic flux and Maxwell stress are out of phase with the gas pressure by $\sim \pi$, and the azimuthal velocity is out of phase with the gas pressure by $\sim \pi/2$.  Furthermore, in geostrophic balance, the gas pressure gradient balances with the Coriolis force,

\begin{equation}
\label{geo}
\frac{\cs^2}{2\Omega}\frac{\partial \overline{\rho}}{\partial x} \approx \rho_0 \overline{\delta v_y}
\end{equation}

\noindent
where the over-bar represents the time and $y,z$ average that is done in Fig.~\ref{st_profile} and $\rho_0$ is the mid-plane gas density.  We have checked this balance by calculating both sides of Equation~(\ref{geo}); we find that as a function of $x$ both sides agree with each other to within a factor of order unity. Thus, the features that we observe in this simulation are indeed zonal flows.

Comparing the location of the magnetic flux bundles with that seen in the snapshot at 30 orbits (Fig.~\ref{beta}), we see that they match exactly.  Clearly, the turbulence seen in our simulations arises from the weakly magnetized, high-density regions of the zonal flows.  This result is consistent with the fact that the MRI requires a weak magnetic field to become active, especially in the presence of strong ambipolar diffusion.

\subsection{Caveat: The ``CWeak" Simulations}
\label{cweak}

Two of our simulations, R100-B4p-CWeak-5.25H and R100-B5p-CWeak-5.25H, which assume cosmic ray ionization at a reduced rate (in addition to $^{26}$Al decay), depict a very laminar magnetic field structure.  Yet, there are clear indications that the gas flow is turbulent (see Fig.~\ref{xe_beta}).  We have analyzed the velocity structure in these simulations and have discovered the presence of an oscillatory, vertically extended circulation pattern, nearly identical to that seen in \cite{gole16}.  While these flows are not turbulent per se, their radial structure is significantly smaller than what can be resolved with ALMA.  Thus, these flows would manifest themselves as turbulent broadening in the observations (Flaherty et al., in prep). 

While the velocity fluctuations seen in these simulations are weak and below (or marginally above) the observational constraints (see Fig.~\ref{xe_beta}), there is a trend of increasing velocities with magnetic field strength.  Thus, we anticipate that for stronger magnetic fields ($\beta_0 \le 10^3$), the velocity fluctuations will increase.  

Upon further examining the magnetic field structure in these simulations, we find that radial and toroidal fields are sitting at the mid-plane in a steady-state configuration.  For such a steady state configuration to exist, the continued shearing of radial field into toroidal field must be balanced by buoyancy and/or diffusion to vertically remove magnetic flux from the mid-plane region.  Such departures from a stationary configuration will likely stir the gas to some extent, inducing fluid motions.  In this particular case, the fluid motions induced are a natural oscillatory mode of the disk \citep{lubow93,gole16}.  That we see stronger gas motions for stronger magnetic fields is consistent with the notion of the magnetic field stirring the gas.  More work is required to fully understand the generation of these modes from the magnetic fields.  However, our preliminary studies suggest that this is a physical effect, as is the increase in velocity fluctuations with magnetic field strength. 
 
\begin{figure*}[ht!]
\begin{center}
\includegraphics[width=1\textwidth,angle=0]{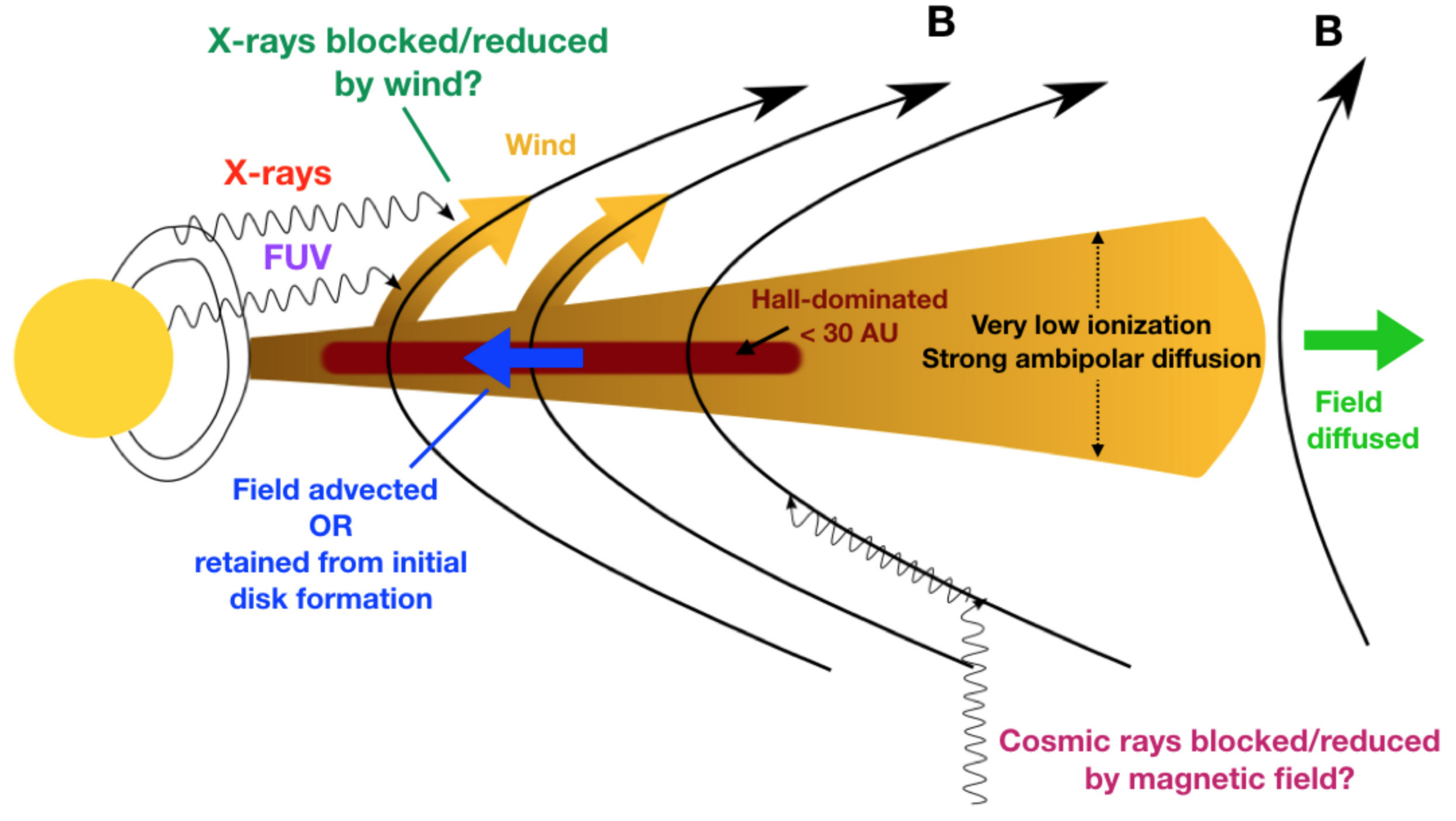}
\end{center}
\caption{Preliminary model for protoplanetary disk structure in systems with weak turbulence in the outer disk. A relatively strong large scale magnetic field exists in the inner disk ($\lesssim 30$AU), either from being advected from larger radii or persisting from the initial formation of the disk.  This field launches a wind, which blocks FUV photons (and possibly X-rays) from reaching the outer disk, while also potentially shielding the outer disk from cosmic rays. This shielding of radiation, coupled with the removal of magnetic flux from the outer disk (either through outward diffusion or inward advection) explains the small turbulent velocities at large distances from the central star.
}
\label{model}
\end{figure*}

\section{Discussion}
\label{discussion}

In the following sections, we discuss the implications of our results for protoplanetary disk structure and evolution.  We first put forth a preliminary model to explain current observations within the context of our simulations, followed by a discussion of support for the model, limitations, and predictions. 

\subsection{A Preliminary Model for Disk Evolution and Structure}

While we have focused here on the HD163926 disk, for which there is significant evidence for low turbulence, there are other systems that display similarly low values (\citealt{flaherty18}, Flaherty et al., in prep).  These results, coupled with the numerical simulations presented here, call into question the notion of an outer disk that is strongly accreting, either through magnetically driven winds or turbulence. Of course, there is ample evidence that gas from the inner disk is accreting onto the star in many of these systems \cite[for HD163926, see][]{mendigutia13}. 

Here, we put forth a preliminary model, described pictorially in Fig.~\ref{model}, to explain current observations that show weak turbulence in the outer disk, but with continued accretion from the inner disk.  A key component in this picture is the presence of a large scale vertical field in the inner regions of the disk ($\lesssim 30$AU).  This vertical field launches a wind (through a Blandford-Payne type process \citealt{blandford82}) that blocks FUV photons (and possibly X-rays; see discussion below) from reaching the outer disk. Furthermore, this magnetic field may also reduce the cosmic ray flux in the outer disk through mechanisms such as magnetic mirroring and funneling (\citealt{cleeves13}, Cleeves, private communication).  A second component to this picture is that any large scale vertical magnetic field present in the outer disk must be relatively weak (though, see discussion below), having been advected inward, or (as supported by some recent global calculations; see \citealt{bai17b}) diffused to very large distances. In either case, most of the magnetic flux remains in the inner disk with less flux present on the scales of $\sim 100$AU.  This model thus satisfies the two conditions we have found necessary for significantly reduced turbulence: low ionization and weak vertical magnetic field in the outer disk.

\subsection{Support for the Model, Limitations, and Caveats}
\label{caveats}

There is both theoretical and observational support for a large scale field in the inner disk.  Recent global simulations of the inner disk region \citep{bai17a} have found no evidence for outward transport of magnetic flux from the inner disk throughout the duration of the simulations, consistent with the maintenance of a strong magnetic flux in the inner disk once it has been developed.  Furthermore, the presence of an inner disk field that launches a wind is consistent with observational constraints found in optical forbidden line studies \citep{msimon16}, and by modeling the near-infrared emission properties of Herbig Ae stars \citep{bans12}.  Specific systems for which there is some evidence of a wind (though, it is possible these winds are photo-evaporative and not magnetic in origin) include V4046Sgr \citep{sacco12}, MWC480 \citep{fernandes18}, TW Hya \citep{pascucci11}, and HD163296 \citep{klaassen13}.  That the two systems currently shown to exhibit weak turbulence also show evidence of a wind provides strong support for our model.

The degree to which ionizing radiation is prevented from reaching the outer disk is more uncertain, however.  In particular, a substantial wind would be required to block X-rays altogether, and indeed, there is observational work \cite[e.g.,][]{pascucci11,pascucci14} suggesting that X-rays can penetrate beyond 1 AU in protoplanetary disks.   It is worth mentioning, however, that while the absorption column for X-rays is quite small (of the order 0.01 g/cm$^2$; \citealt{igea99}), and thus may be blocked by the wind, the scattering component can ionize gas to significantly deeper columns ($\sim$ 1--10 g/cm$^2$ in a protoplanetary disk; \citealt{bai09}).  How precisely this scattering component would penetrate through the wind and impinge upon the disk is not well known, and it may be the case that X-ray flux on the outer disk is significantly reduced as a result of scattering by the wind. Finally, for cosmic rays to be blocked, the magnetic field would likely need to be strongly inclined with respect to the disk rotation axis; if instead the field has a small inclination or is collimated, cosmic rays could reach the outer disk unimpeded. 

However, we reiterate that even if no X-rays and cosmic rays are blocked, one can still achieve turbulent velocities consistent with observations so long as no FUV flux reaches the outer disk {\it and} a sufficiently weak magnetic field is present there (Fig.~\ref{xe_beta}).   When the column required to block FUV is small (e.g., 0.01 g/cm$^2$ \citealt{perez-becker11b}), a modestly strong wind can prevent these photons from reaching the outer disk, as found in some of the global simulations of \cite{bai17a}.   There is observational support for a wind blocking FUV photons as well. For example as argued in \cite{bans12}, a dusty FUV-shielding wind could explain the strong near infrared excess seen in accreting systems around Herbig Ae stars. Indeed, given the large UV flux from the HD163296 star \citep{meeus12}, it is not inconceivable that a substantial dust-filled wind will be required to block UV photons from reaching the outer disk.  

While this preliminary picture is consistent with our observational and theoretical constraints, there remain several uncertainties with this picture and the simulations that inform it. First, as mentioned above, we have neglected the effect of dust on the ionization fraction. While S15 demonstrated that the inclusion of 0.1 $\mu$m grains with mass ratio $10^{-4}$ made very little difference to the amplitude of turbulence in the outer disk compared with the grain-free case\footnote{The ionization fraction is largely unaffected by dust grains when it is higher than the grain abundance \cite[see, e.g.,][]{bai11b}. This is generally the case in the outer disk when the mass fraction of sub-micron grains is $\sim 10^{-4}$ (corresponding to an abundance of $10^{-14}$).}, higher abundance of sub-micron grains could lower the ionization fraction and help suppress turbulence further.  Indeed, such grains may very well be present in protoplanetary disks, at least in the upper layers \citep{bouwman00,bouwman01,sicilia-aguilar07,zsom11}, and if they are sufficiently abundant, the ionization fraction could be driven to such a low value that no turbulence would ensue even in the presence of magnetic fields.  However, the abundance levels of these small grains throughout the vertical extent of the disk is not well-constrained by observations.  Additionally, calculations by \cite{birnstiel11} have demonstrated that in coagulation-fragmentation equilibrium, the number density of grains drops drastically below $\sim 0.1$--$1\mu$m size. These arguments suggest that these small grains may not play a large role in suppressing turbulence. However, the complete exclusion of very small grains remains an assumption that should be explicitly tested; this will be done in a future publication.

Furthermore, as with any study involving local simulations, there remains the question of whether the same behavior would appear in global simulations.  Recent works by \cite{bai17a} and \cite{bethune17} claim their global simulations display mostly laminar fluid motions, which seem to contradict our finding of turbulent motions in local calculations. However, the vast majority of these global simulations were carried out in 2D (or focus on the inner disk, as in \citealt{bai17a}), and it is well-known that allowing for non-axisymmetric behavior in fully 3D simulations can lead to fundamentally different behavior \citep{goodman94,balbus98,simon15b} compared to 2D.  In addition, the one instance of a fully 3D global calculation focused on the outer regions of protoplanetary disks \cite[see][]{bethune17} was carried out at significantly lower effective resolution compared to our local simulations; in these global calculations, the number of grid zones per disk scale height was a factor of 2 (8) lower in radius and height (azimuth) compared to our local models, and the highly diffusive HLLE flux solver \cite[e.g.,][]{oneill12,salvesen14} was used instead of the HLLD solver used in the simulations presented here.  Furthermore, the origin of turbulence, as described in Section~\ref{origin}, is a highly local phenomenon, occurring on scales $\sim H$ as the result of zonal flows. Indeed, zonal flows of comparable scale to those seen here have been seen in global simulations themselves \citep{bethune17}.  It seems likely that in full 3D and with sufficiently high resolution, patches of turbulence will be present in global simulations as well.

The presence of the oscillatory modes as seen in the ``CWeak" simulations (see section~\ref{cweak}), while a physical effect, should also be verified within the context of global simulations.  As pointed out by \cite{gole16}, such modes would likely have somewhat different properties in global disks compared with local models.  If these modes exhibited substantially smaller velocity amplitudes in global simulations, it is possible that the gas velocities in the ``CWeak" simulations would fall below the observational limit regardless of background field strength, which may temper the claim that any large scale magnetic field present in the outer disk should be weak.  However, we suspect that the properties of the modes, particularly their amplitude, would not change significantly in global simulations since the physical mechanism driving the modes would be the same.  In summary, while the numerical results presented here are encouraging, global simulations will be an important next step in testing this preliminary model, and we will pursue such calculations in future studies.

Finally, it is worth commenting on the applicability of our HD163296-based simulations to TW Hya, the only other system currently shown to display weak turbulence \citep{flaherty18}.  Compared to HD163296, the TW Hya disk is lower in mass (\citealt{isella07,bergin13}; F15; \citealt{flaherty18}) and thus has lower gas densities at $R \sim 100$AU, the scales most relevant to ALMA observations (F15; \citealt{flaherty18}).  Since ambipolar diffusion is even more dominant, compared with other non-ideal MHD effects, at lower densities \citep{kunz04}, we expect that the results we present here, which are extracted from ambipolar diffusion dominated simulations, are generally applicable to TW Hya as well as HD163926.  In addition, since TW Hya is nearly face-on, the dominant turbulent velocity to be detected by observations would be the $z$ component. We have examined the turbulent velocity components, namely $xy$ versus $z$ turbulent velocities, in all of our simulations.  The vertical velocity component dominates over (by factors ranging from $2$--$6$) or is very comparable to the $xy$ velocities at heights ranging from $2H$ to $4H$ above the mid-plane.  These results suggest that if present, turbulence would be observable in TW Hya as well.

\subsection{Model Predictions}

Despite the uncertainties discussed in the previous section, our model makes several powerful and testable predictions for disks that show weak turbulence.  First, the outer disk (at scales of $\sim100$AU) should have weak ionization \footnote{This would also be true if there is sufficient dust present to quench all turbulence, as discussed above.}; at the very least, FUV ionization should be absent, and the degree of remaining ionization depends on how much magnetic field exists at these distances (see below).There is already evidence for weak ionization in the ratios of neutral and ionic molecules in the TW Hya disk (\citealt{cleeves15}, but see \citealt{mathews13}), and while recent work \citep{teague16} has put large upper limits on turbulent velocities $\delta v < 0.2$--$0.4\cs$, a new analysis \citep{flaherty18} has suggested that indeed turbulence in this system is quite weak.  

Second, there should be a magnetically launched wind originating from small radii, sufficiently strong so as to largely reduce the FUV (and possibly X-ray) flux at the outer disk. As mentioned above, there are a number of systems that show some evidence of a wind.
 
Finally, any magnetic field present in the outer disk, and thus any wind launched from this region, must be weak.  Conversely, if future observations find a disk with turbulence present, then it will be equally useful to search for high levels of ionization and the presence of magnetic field (e.g., through Zeeman observations) in the outer disk.  Furthermore, these systems should show no sign of a wind launched from the inner disk capable of blocking ionizing radiation. Intriguingly, there is tentative evidence that the system DM Tau is lacking such a wind \citep{msimon16}, which is consistent with preliminary work showing the presence of strong turbulence in this system (Flaherty et al., in prep). These results, if verified, will strongly support our model.

Because of the highly non-uniform distribution of magnetic flux in our model, one further prediction is that accretion is likely very non-steady.  In the case of HD 163296,  the (instantaneous) high accretion rate of $\sim 4.5\times10^{-7} M_\sun/{\rm yr}$ \citep{mendigutia13} is only telling us about the accretion rate through the inner disk; the outer disk can maintain a low accretion rate, consistent with our model.  More support for a non-steady accretion rate can be acquired via a simple timescale argument. Given the disk mass, $0.09 M_\sun$ \citep{isella07}, the instantaneous accretion rate would yield a lifetime of only $\sim$200,000 yr, which is significantly shorter than the actual lifetime ($\sim$ 3 Myr; \citealt{montesinos09}). However, the accretion rate of HD163296 was found to be an order of magnitude lower $\sim$20 years ago \citep{mendigutia13}, (which if more representative of the average accretion rate over the disk's lifetime, would be roughly consistent with the system's age), supporting the notion of a non-steady disk.  Even at this lower accretion rate, however, a significant amount of magnetic flux in the inner disk would be required to drive accretion.  Ultimately, it appears inevitable that in our model, the inner disk will be depleted on a relatively short timescale.  Not only does this predict a non-steady accretion rate, but it may also be related to the formation of transition disks, as some models suggest \citep{wang17}.

Given the trade-off between field strength and ionization level, as shown in Fig.~\ref{xe_beta}, it is difficult to quantify the precise values of the field strength and ionization level required to generate weak turbulence consistent with observations.  However, assuming that $\beta_0 = 10^7$ in the outer disk, then to be consistent with observations, there should be no FUV flux in the outer disk (though X-rays and cosmic rays can still be present).  For the HD163296 disk, this field strength corresponds to $\sim 5$--10$\mu{\rm G}$ at 100 AU, roughly consistent with estimates of Milky Way magnetic field strengths (\citealt{haverkorn15} and references therein).  Further support for such a weak field comes from comparing the ionization levels in our model and the abundance of DCO+ from F17.  The abundance of DCO+ puts a lower limit on the ionization fraction at the disk mid-plane of $\sim 10^{-11}$. This is consistent with the ionization level in our simulations that include X-ray and cosmic ray ionization (at the standard ionization rate of $10^{-17} {\rm s}^{-1}$). However, for the ``weak CR" simulations, the ionization fraction is an order of magnitude lower than this, suggesting that in HD163296, X-rays and cosmic rays are not blocked, whereas FUV photons are, and there is a weak vertical magnetic field at large distances from the star.

Of course, if the overall global magnetic field is stronger ($\beta_0 < 10^7$), then one might expect that this stronger magnetic field would allow less ionizing flux from reaching the outer disk, possibly blocking X-ray photons in addition to FUV. This would be consistent with the simulation at $\beta_0 = 10^5$ with both FUV and X-ray ionization removed.  While potentially not applicable to HD163296, as discussed above, this picture may apply to other disk systems. Ultimately, however, more sophisticated fully three-dimensional global calculations will be required to determine the magnetic field strength needed to block ionizing sources of radiation.

\section{Conclusions}
\label{conclusions}

We have carried out a series of local, shearing box simulations in order to further quantity the influence of magnetic fields on protoplanetary disks in the presence of low ionization.  Our primary results are as follows:

\begin{itemize}

\item For a relatively strong background vertical field ($\beta_0 = 10^3$), which drives a largely laminar wind, turbulence is still present at values that are inconsistent with observational constraints and arises from regions of zonal flows where the vertical magnetic field is sufficiently weak for the MRI to persist.  Magnetic winds are not the solution to the weak turbulence issue. 

\item Only with a weak vertical magnetic field {\it and} a low ionization fraction are turbulent velocities consistent with observations.  Furthermore, a trade-off exists; if the magnetic field strength (ionization fraction) is increased, then
the ionization fraction (magnetic field strength) must be decreased to match observations.

\item A preliminary picture to explain these observational and theoretical constraints is that a large-scale vertical magnetic field must be present in the inner disk to launch a strong wind that shields the outer disk from ionizing radiation, and the outer disk
itself has a very weak (or no) large scale field.

\item Such a non-uniform distribution of magnetic flux implies that such disks are non-steady, which, in the case of HD163296, is supported by observational measurements of the accretion rate onto the central star.

\end{itemize}

While the details of this preliminary model need to be tested with fully 3D global MHD simulations, the model predicts that
if indeed there is low turbulence inferred in the outer disk, then observations should constrain the ionization rate in the outer disk to be relatively small and should demonstrate the presence of a large-scale magnetically launched wind in the inner disk.  With ALMA operations continuing well into the coming decade, we should have many opportunities to test these predictions.

\acknowledgments
We thank Phil Armitage, Ilse Cleeves, Ilaria Pascucci, Richard Teague, and Upasana Das for useful discussions related to this work.  We also thank the anonymous referee whose comments improved the quality of this paper. 
JBS acknowledges support from  NASA grants NNX13AI58G and NNX16AB42G, and from NSF grant AST 1313021. The computations were performed on {\sc Stampede}, {\sc Stampede 2}, and {\sc Maverick} through XSEDE grants TG-AST120062 and TG-AST140001. We thank the Kavli Institute for Theoretical Physics, supported in part by the National Science Foundation under Grant No. NSF PHY-1125915, for hospitality during the completion of the paper. 

\software{{\sc Athena} \citep{stone08}}


\begin{thebibliography}{40}
\expandafter\ifx\csname natexlab\endcsname\relax\def\natexlab#1{#1}\fi

\bibitem[{Bai(2011)}]{bai11b}
Bai, X.-N. 2011, The Astrophysical Journal, 739, 50

\bibitem[{Bai(2013)}]{bai13c}
---. 2013, The Astrophysical Journal, 772, 96

\bibitem[{Bai(2014)}]{bai14a}
---. 2014, The Astrophysical Journal, 791, 137

\bibitem[{Bai(2015)}]{bai15}
---. 2015, The Astrophysical Journal, 798, 84

\bibitem[{{Bai}(2017)}]{bai17}
{Bai}, X.-N. 2017, \apj, 845, 75

\bibitem[{Bai \& Stone(2011)}]{bai11a}
Bai, X.-N., \& Stone, J.~M. 2011, The Astrophysical Journal, 736, 144

\bibitem[{Bai \& Stone(2013)}]{bai13b}
---. 2013, The Astrophysical Journal, 769, 76

\bibitem[{Balbus \& Hawley(1998)}]{balbus98}
Balbus, S.~A., \& Hawley, J.~F. 1998, Reviews of Modern Physics, 70, 1

\bibitem[{Baruteau {et~al.}(2011)Baruteau, Fromang, Nelson, \&
  Masset}]{baruteau11}
Baruteau, C., Fromang, S., Nelson, R.~P., \& Masset, F. 2011, Astronomy and
  Astrophysics, 533, A84

\bibitem[{{B{\'e}thune} {et~al.}(2017){B{\'e}thune}, {Lesur}, \&
  {Ferreira}}]{bethune17}
{B{\'e}thune}, W., {Lesur}, G., \& {Ferreira}, J. 2017, \aap, 600, A75

\bibitem[{Birnstiel {et~al.}(2010)Birnstiel, Dullemond, \&
  Brauer}]{birnstiel10}
Birnstiel, T., Dullemond, C.~P., \& Brauer, F. 2010, Astronomy and Astrophysics

\bibitem[{Blandford \& Payne(1982)}]{blandford82}
Blandford, R.~D., \& Payne, D.~G. 1982, Monthly Notices of the Royal
  Astronomical Society, 199, 883

\bibitem[{Clarke \& Pringle(1988)}]{clarke88}
Clarke, C.~J., \& Pringle, J.~E. 1988, Monthly Notices of the Royal
  Astronomical Society (ISSN 0035-8711), 235, 365

\bibitem[{Cuzzi {et~al.}(2008)Cuzzi, Hogan, \& Shariff}]{cuzzi08}
Cuzzi, J.~N., Hogan, R.~C., \& Shariff, K. 2008, The Astrophysical Journal,
  687, 1432

\bibitem[{Flaherty {et~al.}(2015)Flaherty, Hughes, Rosenfeld, Andrews, Chiang,
  Simon, Kerzner, \& Wilner}]{flaherty15}
Flaherty, K.~M., Hughes, A.~M., Rosenfeld, K.~A., {et~al.} 2015, The
  Astrophysical Journal, 813, 99

\bibitem[{Flaherty {et~al.}(2017)Flaherty, Hughes, Rose, Simon, Qi, Andrews,
  K{\'o}sp{\'a}l, Wilner, Chiang, Armitage, \& Bai}]{flaherty17}
Flaherty, K.~M., Hughes, A.~M., Rose, S.~C., {et~al.} 2017, The Astrophysical
  Journal, 843, 150

\bibitem[{Fromang \& Papaloizou(2006)}]{fromang06a}
Fromang, S., \& Papaloizou, J. 2006, A\&A, 452, 751

\bibitem[{Gammie(1996)}]{gammie96}
Gammie, C.~F. 1996, ApJ, 457, 355

\bibitem[{Guilloteau {et~al.}(2012)Guilloteau, Dutrey, Wakelam, Hersant,
  Semenov, Chapillon, Henning, \& Pi{\'e}tu}]{guilloteau12}
Guilloteau, S., Dutrey, A., Wakelam, V., {et~al.} 2012, Astronomy and
  Astrophysics, 548, 70

\bibitem[{Hawley {et~al.}(1995)Hawley, Gammie, \& Balbus}]{hawley95a}
Hawley, J.~F., Gammie, C.~F., \& Balbus, S.~A. 1995, ApJ, 440, 742

\bibitem[{Johansen {et~al.}(2009)Johansen, Youdin, \& Klahr}]{johansen09a}
Johansen, A., Youdin, A., \& Klahr, H. 2009, The Astrophysical Journal, 697,
  1269

\bibitem[{Kunz(2008)}]{kunz08}
Kunz, M.~W. 2008, Monthly Notices of the Royal Astronomical Society, 385, 1494

\bibitem[{Kunz \& Lesur(2013)}]{kunz13}
Kunz, M.~W., \& Lesur, G. 2013, Monthly Notices of the Royal Astronomical
  Society, 434, 2295

\bibitem[{Lesur {et~al.}(2014)Lesur, Kunz, \& Fromang}]{lesur14}
Lesur, G., Kunz, M.~W., \& Fromang, S. 2014, Astronomy and Astrophysics, 566,
  56

\bibitem[{Lubow \& Ida(2011)}]{lubow11}
Lubow, S.~H., \& Ida, S. 2011, Exoplanets, 347

\bibitem[{Nelson \& Papaloizou(2004)}]{nelson04}
Nelson, R.~P., \& Papaloizou, J. C.~B. 2004, MNRAS, 350, 849

\bibitem[{Paardekooper {et~al.}(2011)Paardekooper, Baruteau, \&
  Kley}]{paardekooper11}
Paardekooper, S.~J., Baruteau, C., \& Kley, W. 2011, Monthly Notices of the
  Royal Astronomical Society, 410, 293

\bibitem[{Perez-Becker \& Chiang(2011)}]{perez-becker11b}
Perez-Becker, D., \& Chiang, E. 2011, The Astrophysical Journal, 735, 8

\bibitem[{Salmeron {et~al.}(2007)Salmeron, K{\"o}nigl, \& Wardle}]{salmeron07}
Salmeron, R., K{\"o}nigl, A., \& Wardle, M. 2007, MNRAS, 375, 177

\bibitem[{Simon \& Armitage(2014)}]{simon14}
Simon, J.~B., \& Armitage, P.~J. 2014, The Astrophysical Journal, 784, 15

\bibitem[{Simon {et~al.}(2011)Simon, Armitage, \& Beckwith}]{simon11b}
Simon, J.~B., Armitage, P.~J., \& Beckwith, K. 2011, The Astrophysical Journal,
  743, 17

\bibitem[{Simon {et~al.}(2013b)Simon, Bai, Armitage, Stone, \&
  Beckwith}]{simon13b}
Simon, J.~B., Bai, X.-N., Armitage, P.~J., Stone, J.~M., \& Beckwith, K. 2013b,
  The Astrophysical Journal, 775, 73

\bibitem[{Simon {et~al.}(2013a)Simon, Bai, Stone, Armitage, \&
  Beckwith}]{simon13a}
Simon, J.~B., Bai, X.-N., Stone, J.~M., Armitage, P.~J., \& Beckwith, K. 2013a,
  The Astrophysical Journal, 764, 66

\bibitem[{Simon {et~al.}(2015{\natexlab{a}})Simon, Hughes, Flaherty, Bai, \&
  Armitage}]{simon15a}
Simon, J.~B., Hughes, A.~M., Flaherty, K.~M., Bai, X.-N., \& Armitage, P.~J.
  2015{\natexlab{a}}, The Astrophysical Journal, 808, 180

\bibitem[{Simon {et~al.}(2015{\natexlab{b}})Simon, Lesur, Kunz, \&
  Armitage}]{simon15b}
Simon, J.~B., Lesur, G., Kunz, M.~W., \& Armitage, P.~J. 2015{\natexlab{b}},
  Monthly Notices of the Royal Astronomical Society, 454, 1117

\bibitem[{Stone {et~al.}(2008)Stone, Gardiner, Teuben, Hawley, \&
  Simon}]{stone08}
Stone, J.~M., Gardiner, T.~A., Teuben, P., Hawley, J.~F., \& Simon, J.~B. 2008,
  The Astrophysical Journal Supplement, 178, 137

\bibitem[{Suzuki \& Inutsuka(2009)}]{suzuki09}
Suzuki, T.~K., \& Inutsuka, S.-I. 2009, ApJ, 691, L49

\bibitem[{Teague {et~al.}(2016)Teague, Guilloteau, \& Semenov}]{teague16}
Teague, R., Guilloteau, S., \& Semenov, D. 2016, Astronomy {\&} {\ldots}

\bibitem[{van~den Ancker {et~al.}(1998)van~den Ancker, de~Winter, \& Tjin
  A~Djie}]{vandenancker98}
van~den Ancker, M.~E., de~Winter, D., \& Tjin A~Djie, H. R.~E. 1998, Astronomy
  and Astrophysics, 330, 145

\bibitem[{Youdin \& Lithwick(2007)}]{youdin07b}
Youdin, A.~N., \& Lithwick, Y. 2007, Icarus, Icarus, 588

\end{thebibliography}


\begin{thebibliography}{40}
\expandafter\ifx\csname natexlab\endcsname\relax\def\natexlab#1{#1}\fi

\bibitem[Bai \& Goodman(2009)]{bai09} Bai, X.-N., \& Goodman, J.\ 2009, \apj, 701, 737 

\bibitem[{Bai(2011)}]{bai11b}
Bai, X.-N. 2011, The Astrophysical Journal, 739, 50

\bibitem[{Bai(2013)}]{bai13c}
Bai, X.-N. 2013, The Astrophysical Journal, 772, 96

\bibitem[{Bai(2014)}]{bai14a}
Bai, X.-N. 2014, The Astrophysical Journal, 791, 137

\bibitem[Bai \& Stone(2014)]{bai14b} Bai, X.-N., \& Stone, J.~M.\ 2014, \apj, 796, 31 


\bibitem[{Bai(2015)}]{bai15}
Bai, X.-N. 2015, The Astrophysical Journal, 798, 84

\bibitem[{{Bai}(2017)}]{bai17a}
{Bai}, X.-N. 2017, \apj, 845, 75

\bibitem[{Bai \& Stone(2011)}]{bai11a}
Bai, X.-N., \& Stone, J.~M. 2011, The Astrophysical Journal, 736, 144

\bibitem[{Bai \& Stone(2013)}]{bai13b}
Bai, X.-N., \& Stone, J.~M. 2013, The Astrophysical Journal, 769, 76

\bibitem[Bai \& Stone(2017)]{bai17b} Bai, X.-N., \& Stone, J.~M.\ 2017, \apj, 836, 46 

\bibitem[{Balbus \& Hawley(1998)}]{balbus98}
Balbus, S.~A., \& Hawley, J.~F. 1998, Reviews of Modern Physics, 70, 1

\bibitem[Bans \& K{\"o}nigl(2012)]{bans12} Bans, A., \& K{\"o}nigl, A.\ 2012, \apj, 758, 100

\bibitem[{Baruteau {et~al.}(2011)Baruteau, Fromang, Nelson, \&
  Masset}]{baruteau11}
Baruteau, C., Fromang, S., Nelson, R.~P., \& Masset, F. 2011, Astronomy and
  Astrophysics, 533, A84
  
 \bibitem[Bergin et al.(2013)]{bergin13} Bergin, E.~A., Cleeves, L.~I., Gorti, U., et al.\ 2013, \nat, 493, 644 

\bibitem[{{B{\'e}thune} {et~al.}(2017){B{\'e}thune}, {Lesur}, \&
  {Ferreira}}]{bethune17}
{B{\'e}thune}, W., {Lesur}, G., \& {Ferreira}, J. 2017, \aap, 600, A75

\bibitem[{Birnstiel {et~al.}(2010)Birnstiel, Dullemond, \&
  Brauer}]{birnstiel10}
Birnstiel, T., Dullemond, C.~P., \& Brauer, F. 2010, Astronomy and Astrophysics

\bibitem[Birnstiel et al.(2011)]{birnstiel11} Birnstiel, T., Ormel, C.~W., \& Dullemond, C.~P.\ 2011, \aap, 525, A11 

\bibitem[{Blandford \& Payne(1982)}]{blandford82}
Blandford, R.~D., \& Payne, D.~G. 1982, Monthly Notices of the Royal
  Astronomical Society, 199, 883
  
\bibitem[Bouwman et al.(2000)]{bouwman00} Bouwman, J., de Koter, A., van den Ancker, M.~E., \& Waters, L.~B.~F.~M.\ 2000, \aap, 360, 213
  
\bibitem[Bouwman et al.(2001)]{bouwman01} Bouwman, J., Meeus, G., de Koter, A., et al.\ 2001, \aap, 375, 950 

\bibitem[Cleeves et al.(2013)]{cleeves13} Cleeves, L.~I., Adams, F.~C., \& Bergin, E.~A.\ 2013, \apj, 772, 5 
  
\bibitem[Cleeves et al.(2015)]{cleeves15} Cleeves, L.~I., Bergin, E.~A., Qi, C., Adams, F.~C., \& {\"O}berg, K.~I.\ 2015, \apj, 799, 204 

\bibitem[{Cuzzi {et~al.}(2008)Cuzzi, Hogan, \& Shariff}]{cuzzi08}
Cuzzi, J.~N., Hogan, R.~C., \& Shariff, K. 2008, The Astrophysical Journal,
  687, 1432
  
 \bibitem[Fernandes et al.(2018)]{fernandes18} Fernandes, R.~B., Long, Z.~C., Pikhartova, M., et al.\ 2018, \apj, 856, 103

\bibitem[{Flaherty {et~al.}(2015)Flaherty, Hughes, Rosenfeld, Andrews, Chiang,
  Simon, Kerzner, \& Wilner}]{flaherty15}
Flaherty, K.~M., Hughes, A.~M., Rosenfeld, K.~A., {et~al.} 2015, The
  Astrophysical Journal, 813, 99

\bibitem[{Flaherty {et~al.}(2017)Flaherty, Hughes, Rose, Simon, Qi, Andrews,
  K{\'o}sp{\'a}l, Wilner, Chiang, Armitage, \& Bai}]{flaherty17}
Flaherty, K.~M., Hughes, A.~M., Rose, S.~C., {et~al.} 2017, The Astrophysical
  Journal, 843, 150
  
\bibitem[Flaherty et al.(2018)]{flaherty18} Flaherty, K.~M., Hughes, A.~M., Teague, R., et al.\ 2018, \apj, 856, 117 

\bibitem[{Fromang \& Papaloizou(2006)}]{fromang06a}
Fromang, S., \& Papaloizou, J. 2006, A\&A, 452, 751

\bibitem[{Gammie(1996)}]{gammie96}
Gammie, C.~F. 1996, ApJ, 457, 355

\bibitem[Gole et al.(2016)]{gole16} Gole, D., Simon, J.~B., Lubow, S.~H., \& Armitage, P.~J.\ 2016, \apj, 826, 18 

\bibitem[Goodman \& Xu(1994)]{goodman94} Goodman, J., \& Xu, G.\ 1994, \apj, 432, 213

\bibitem[{Guilloteau {et~al.}(2012)Guilloteau, Dutrey, Wakelam, Hersant,
  Semenov, Chapillon, Henning, \& Pi{\'e}tu}]{guilloteau12}
Guilloteau, S., Dutrey, A., Wakelam, V., {et~al.} 2012, Astronomy and
  Astrophysics, 548, 70
  
  \bibitem[{G{\"u}nther \& Schmitt(2009)}]{gunther09}
G{\"u}nther, H.~M., \& Schmitt, J. H. M.~M. 2009, Astronomy and Astrophysics,
  494, 1041

\bibitem[Haverkorn(2015)]{haverkorn15} Haverkorn, M.\ 2015, Magnetic Fields in Diffuse Media, 407, 483 

\bibitem[{Hawley {et~al.}(1995)Hawley, Gammie, \& Balbus}]{hawley95a}
Hawley, J.~F., Gammie, C.~F., \& Balbus, S.~A. 1995, ApJ, 440, 742

\bibitem[Igea \& Glassgold(1999)]{igea99} Igea, J., \& Glassgold, A.~E.\ 1999, \apj, 518, 848 


\bibitem[Isella et al.(2007)]{isella07} Isella, A., Testi, L., Natta, A., et al.\ 2007, \aap, 469, 213 

\bibitem[{Johansen {et~al.}(2009)Johansen, Youdin, \& Klahr}]{johansen09a}
Johansen, A., Youdin, A., \& Klahr, H. 2009, The Astrophysical Journal, 697,
  1269

\bibitem[Klaassen et al.(2013)]{klaassen13} Klaassen, P.~D., Juhasz, A., Mathews, G.~S., et al.\ 2013, \aap, 555, A73 

\bibitem[Kunz \& Balbus(2004)]{kunz04} Kunz, M.~W., \& Balbus, S.~A.\ 2004, \mnras, 348, 355 

\bibitem[{Kunz(2008)}]{kunz08}
Kunz, M.~W. 2008, Monthly Notices of the Royal Astronomical Society, 385, 1494

\bibitem[{Kunz \& Lesur(2013)}]{kunz13}
Kunz, M.~W., \& Lesur, G. 2013, Monthly Notices of the Royal Astronomical
  Society, 434, 2295

\bibitem[{Lesur {et~al.}(2014)Lesur, Kunz, \& Fromang}]{lesur14}
Lesur, G., Kunz, M.~W., \& Fromang, S. 2014, Astronomy and Astrophysics, 566,
  56

\bibitem[{Lubow \& Ida(2011)}]{lubow11}
Lubow, S.~H., \& Ida, S. 2011, Exoplanets, 347

\bibitem[Lubow \& Pringle(1993)]{lubow93} Lubow, S.~H., \& Pringle, J.~E.\ 1993, \apj, 409, 360

\bibitem[Mathews et al.(2013)]{mathews13} Mathews, G.~S., Klaassen, P.~D., Juh{\'a}sz, A., et al.\ 2013, \aap, 557, A132 

\bibitem[Meeus et al.(2012)]{meeus12} Meeus, G., Montesinos, B., Mendigut{\'{\i}}a, I., et al.\ 2012, \aap, 544, A78 

\bibitem[Mendigut{\'{\i}}a et al.(2013)]{mendigutia13} Mendigut{\'{\i}}a, I., Brittain, S., Eiroa, C., et al.\ 2013, \apj, 776, 44 

\bibitem[Montesinos et al.(2009)]{montesinos09} Montesinos, B., Eiroa, C., Mora, A., \& Mer{\'{\i}}n, B.\ 2009, \aap, 495, 901 

\bibitem[{Nelson \& Papaloizou(2004)}]{nelson04}
Nelson, R.~P., \& Papaloizou, J. C.~B. 2004, MNRAS, 350, 849

\bibitem[Okuzumi et al.(2016)]{okuzumi16} Okuzumi, S., Momose, M., Sirono, S.-i., Kobayashi, H., \& Tanaka, H.\ 2016, \apj, 821, 82 

\bibitem[O'Neill et al.(2012)]{oneill12} O'Neill, S.~M., Beckwith, K., \& Begelman, M.~C.\ 2012, \mnras, 422, 1436 

\bibitem[{Paardekooper {et~al.}(2011)Paardekooper, Baruteau, \&
  Kley}]{paardekooper11}
Paardekooper, S.~J., Baruteau, C., \& Kley, W. 2011, Monthly Notices of the
  Royal Astronomical Society, 410, 293
  
  \bibitem[Pandey \& Wardle(2012)]{pandey12} Pandey, B.~P., \& Wardle, M.\ 2012, \mnras, 423, 222 
  
 \bibitem[Pascucci et al.(2011)]{pascucci11} Pascucci, I., Sterzik, M., Alexander, R.~D., et al.\ 2011, \apj, 736, 13  
 
 \bibitem[Pascucci et al.(2014)]{pascucci14} Pascucci, I., Ricci, L., Gorti, U., et al.\ 2014, \apj, 795, 1 
  
\bibitem[{Perez-Becker \& Chiang(2011)}]{perez-becker11b}
Perez-Becker, D., \& Chiang, E. 2011, The Astrophysical Journal, 735, 8

\bibitem[Pinte et al.(2016)]{pinte16} Pinte, C., Dent, W.~R.~F., M{\'e}nard, F., et al.\ 2016, \apj, 816, 25 

\bibitem[Rosenfeld et al.(2013)]{rosenfeld13} Rosenfeld, K.~A., Andrews, S.~M., Hughes, A.~M., Wilner, D.~J., \& Qi, C.\ 2013, \apj, 774, 16 

\bibitem[Sacco et al.(2012)]{sacco12} Sacco, G.~G., Flaccomio, E., Pascucci, I., et al.\ 2012, \apj, 747, 142 

\bibitem[{Salmeron {et~al.}(2007)Salmeron, K{\"o}nigl, \& Wardle}]{salmeron07}
Salmeron, R., K{\"o}nigl, A., \& Wardle, M. 2007, MNRAS, 375, 177

\bibitem[Salvesen et al.(2014)]{salvesen14} Salvesen, G., Beckwith, K., Simon, J.~B., O'Neill, S.~M., \& Begelman, M.~C.\ 2014, \mnras, 438, 1355

\bibitem[{Shakura \& Sunyaev(1973)}]{shakura73}
Shakura, N.~I., \& Syunyaev, R.~A. 1973, A\&A, 24, 337

\bibitem[Sicilia-Aguilar et al.(2007)]{sicilia-aguilar07} Sicilia-Aguilar, A., Hartmann, L.~W., Watson, D., et al.\ 2007, \apj, 659, 1637 

\bibitem[{Simon \& Armitage(2014)}]{simon14}
Simon, J.~B., \& Armitage, P.~J. 2014, The Astrophysical Journal, 784, 15

\bibitem[{Simon {et~al.}(2011)Simon, Armitage, \& Beckwith}]{simon11b}
Simon, J.~B., Armitage, P.~J., \& Beckwith, K. 2011, The Astrophysical Journal,
  743, 17

\bibitem[{Simon {et~al.}(2013b)Simon, Bai, Armitage, Stone, \&
  Beckwith}]{simon13b}
Simon, J.~B., Bai, X.-N., Armitage, P.~J., Stone, J.~M., \& Beckwith, K. 2013b,
  The Astrophysical Journal, 775, 73

\bibitem[{Simon {et~al.}(2013a)Simon, Bai, Stone, Armitage, \&
  Beckwith}]{simon13a}
Simon, J.~B., Bai, X.-N., Stone, J.~M., Armitage, P.~J., \& Beckwith, K. 2013a,
  The Astrophysical Journal, 764, 66

\bibitem[{Simon {et~al.}(2015{\natexlab{a}})Simon, Hughes, Flaherty, Bai, \&
  Armitage}]{simon15a}
Simon, J.~B., Hughes, A.~M., Flaherty, K.~M., Bai, X.-N., \& Armitage, P.~J.
  2015{\natexlab{a}}, The Astrophysical Journal, 808, 180

\bibitem[{Simon {et~al.}(2015{\natexlab{b}})Simon, Lesur, Kunz, \&
  Armitage}]{simon15b}
Simon, J.~B., Lesur, G., Kunz, M.~W., \& Armitage, P.~J. 2015{\natexlab{b}},
  Monthly Notices of the Royal Astronomical Society, 454, 1117

\bibitem[Simon et al.(2016)]{msimon16} Simon, M.~N., Pascucci, I., Edwards, S., et al.\ 2016, \apj, 831, 169 

\bibitem[{Stone {et~al.}(2008)Stone, Gardiner, Teuben, Hawley, \&
  Simon}]{stone08}
Stone, J.~M., Gardiner, T.~A., Teuben, P., Hawley, J.~F., \& Simon, J.~B. 2008,
  The Astrophysical Journal Supplement, 178, 137

\bibitem[{Suzuki \& Inutsuka(2009)}]{suzuki09}
Suzuki, T.~K., \& Inutsuka, S.-I. 2009, ApJ, 691, L49

\bibitem[{Swartz {et~al.}(2005)Swartz, Drake, Elsner, Ghosh, Grady, Wassell,
  Woodgate, \& Kimble}]{swartz05}
Swartz, D.~A., Drake, J.~J., Elsner, R.~F., {et~al.} 2005, The Astrophysical
  Journal, 628, 811

\bibitem[{Teague {et~al.}(2016)Teague, Guilloteau, \& Semenov}]{teague16}
Teague, R., Guilloteau, S., \& Semenov, D. 2016, Astronomy {\&} {\ldots}

\bibitem[{{Umebayashi} \& {Nakano}(1981)}]{umebayashi81}
{Umebayashi}, T., \& {Nakano}, T. 1981, \pasj, 33, 617

\bibitem[{van~den Ancker {et~al.}(1998)van~den Ancker, de~Winter, \& Tjin
  A~Djie}]{vandenancker98}
van~den Ancker, M.~E., de~Winter, D., \& Tjin A~Djie, H. R.~E. 1998, Astronomy
  and Astrophysics, 330, 145
  
\bibitem[Wang \& Goodman(2017)]{wang17} Wang, L., \& Goodman, J.~J.\ 2017, \apj, 835, 59

\bibitem[{Youdin \& Lithwick(2007)}]{youdin07b}
Youdin, A.~N., \& Lithwick, Y. 2007, Icarus, Icarus, 588

\bibitem[Zsom et al.(2011)]{zsom11} Zsom, A., Ormel, C.~W., Dullemond, C.~P., \& Henning, T.\ 2011, \aap, 534, A73 

\end{thebibliography}
\end{document}